\documentclass[aps,twocolumn,showpacs,nofootinbib]{revtex4-1}

\usepackage{amsmath,amssymb,bm}
\usepackage[dvips]{graphicx}
\usepackage{hyperref}
\usepackage{yfonts}
\usepackage{color}
\newcommand{\beq}{\begin{eqnarray}}
\newcommand{\eeq}{\end{eqnarray}}
\renewcommand\d{\partial}
\begin{document}

\title{Axion electrodynamics and nonrelativistic photons 
in nuclear and quark matter}

\author{Naoki Yamamoto}
\affiliation{Department of Physics, Keio University, Yokohama 223-8522, Japan}

\begin{abstract}
We argue that the effective theory for electromagnetic fields in spatially varying meson 
condensations in dense nuclear and quark matter is given by the axion electrodynamics. 
We show that one of the helicity states of photons there has the nonrelativistic gapless 
dispersion relation $\omega \sim k^2$ at small momentum, while the other is gapped. 
This ``nonrelativistic photon" may also be realized at the interface between topological 
and trivial insulators in condensed matter systems.
\end{abstract}
%\pacs{
% 12.38.Aw General properties of QCD
% 73.43.-f	 Quantum Hall effects
% 14.80.Va Axions and other Nambu-Goldstone bosons (Majorons, familons, etc.)
% 26.60.-c Nuclear matter aspects of neutron stars}
\maketitle

\section{Introduction} 
Recently, topologically nontrivial states of matter, such as the topological insulators 
and topological superconductors \cite{Hasan:2010xy, Qi:2011zya}, have attracted much 
attention. The nontrivial topology of the fully gaped bulk is reflected in the physics of 
boundaries. The well-known example is the boundary between a topological insulator 
and an ordinary (or topologically trivial) insulator \cite{Qi:2008ew}, where electromagnetic 
responses are modified and are described by the axion electrodynamics \cite{Wilczek:1987mv}. 

In this paper, we point out yet another realization of the essentially same physics of this 
surface state in a different system: spatially varying meson condensations in dense nuclear 
and quark matter, characterized by a nonzero gradient of some meson field (say $\varphi$), 
$\langle {\bm \nabla} \varphi \rangle \neq 0$. Such phases appear in nuclear matter in a 
strong magnetic field (known as the pion domain wall) \cite{Son:2007ny} and in color 
superconducting quark matter even in the absence of the magnetic field (known as the 
meson supercurrent phase) \cite{Gerhold:2006dt, Gerhold:2006np}, which may be realized 
inside neutron stars or quark stars.

We argue, based only on the symmetries and topology of the system, that the low-energy 
dynamics of electromagnetic fields there is described by the axion electrodynamics. 
Similarly to the boundary of the topological insulator, the anomalous magnetoelectric effects 
(e.g., the quantum Hall effect) appear in these phases.

We also show that, due to the modifications of the electromagnetic dynamics, photons 
behave differently depending on the helicity at small momentum; one of the helicity states 
has the nonrelativistic gapless dispersion relation $\omega \sim k^2$, while the other is 
gapped [see Eq.~(\ref{dispersion_long})]. Such behavior, including the ``nonrelativistic 
gapless photon," is a universal feature of the axion electrodynamics not only there, but also 
at the interface between topological and ordinary insulators. To the extent of our knowledge, 
this is a new result, even in the latter context, which may be tested in table-top experiments.

In this paper, we take the units $\hbar = c = e = 1$.

\section{Spatially varying meson condensations}
We first review the possible spatially varying meson condensations in high density matter 
\cite{Son:2007ny, Gerhold:2006dt, Gerhold:2006np}.
% (see also Ref.~\cite{Thompson:2008qw} for its holographic realization).
These examples in an external magnetic field include the $\pi^0$ domain wall in nuclear matter, 
the $\eta$ meson domain wall in two-flavor color superconductivity (2SC), and the $\eta'$ 
meson domain wall in color-flavor locked (CFL) phase \cite{Son:2007ny}. The space-dependent
meson condensation may also appear \emph{even without} the external magnetic field at 
intermediate density between 2SC and CFL phases due the Fermi surface splitting induced by 
the large strange quark mass ($m_{\rm s} \gg m_{\rm u,d}$); in this context, it is called the 
meson supercurrent phase \cite{Gerhold:2006dt, Gerhold:2006np}. 

Our argument below is applicable to all of them, but for simplicity and concreteness, 
we will consider a space-dependent $\pi^0$ field (in nuclear matter) as an example. Applications 
of our argument to other possible spatially varying meson condensations are straightforward.

At low energy, the dynamics of QCD can be systematically described by the effective 
theory---the chiral perturbation theory for pions. Let us concentrate on the dynamics of 
neutral pion (which we denote by $\pi^0$), as the other degrees of freedom ($\pi^{\pm}$) 
will be irrelevant in our discussion. The effective Lagrangian is given by
\beq
\label{L}
{\cal L}_{\pi^0} = \frac{1}{2}(\d_{\mu} \pi^0)^2 + m_{\pi}^2 f_{\pi}^2 \cos \left(\frac{\pi^0}{f_{\pi}} \right)\,,
\eeq
where $f_{\pi}$ the pion decay constant and $m_{\pi}$ is the pion mass.

The equation of motion for $\phi \equiv \pi^0/f_{\pi}$ derived from Eq.~(\ref{L}) is 
given by the sine-Gordon equation,
\beq
\label{SG}
\d_t^2 \phi - \d_z^2 \phi + m_{\pi}^2 \sin \phi =0,
\eeq
where, for simplicity, we assume $\phi$ depends only on $z$. 
This has the domain-wall solution,
\beq
\label{DW}
\phi (z) = 4 \tan^{-1} e^{m_{\pi} z}.
\eeq
From this solution, it is easy to see that $\phi$ changes $2\pi$ across the wall and the 
thickness of the wall is $\sim 1/m_{\pi}$. The energy density per unit area is 
\beq
\label{E}
\frac{E}{S} = 8 f_{\pi}^2 m_{\pi}.
\eeq

At finite baryon chemical potential $\mu_{\rm B}$ in an external magnetic field, we have 
the additional  ``Wess-Zumino-Witten (WZW) term" related to quantum anomalies in the 
effective Lagrangian \cite{Son:2007ny},
\beq
\label{WZW}
{\cal L}_{B} = C \mu_{\rm B} {\bm B}_{\rm ex} \cdot {\bm \nabla} \pi^0,
\eeq
where 
\beq
\label{C}
C = \frac{1}{4\pi^2 f_{\pi}}\,.
\eeq
This term gives the domain wall a nonzero baryon density given by
\beq
n_{\rm B} = C {\bm B}_{\rm ex} \cdot {\bm \nabla} \pi^0\,,
\eeq
so that baryon number per unit surface area is
\beq
\label{N}
\frac{N_{\rm B}}{S} = \frac{B_{\rm ex}}{2\pi}\,,
\eeq
where $B_{\rm ex} \equiv |{\bm B}_{\rm ex}|$. 

Combining Eqs.~(\ref{E}) and (\ref{N}), the energy of the domain wall per baryon number is 
\beq
\frac{E}{N_{\rm B}} = \frac{16 \pi f_{\pi}^2 m_{\pi}}{B_{\rm ex}}\,.
\eeq
When this energy becomes smaller than the baryon mass $m_{\rm N}$, namely,
\beq
\label{bound}
B_{\rm ex} > \frac{16 \pi f_{\pi}^2 m_{\pi}}{m_{\rm N}}\,,
\eeq
the $\pi^0$ domain wall is energetically more favorable than nuclear matter \cite{Son:2007ny}; 
the ground state becomes the $\pi^0$ domain wall. A more elaborate analysis shows that the 
most energetically favorable ground state is the periodic array of $\pi^0$ domain walls that 
spontaneously break parity and continuous translational symmetries \cite{CMSL}. The structure
of this state is mathematically similar to the ``chiral magnetic soliton lattice" experimentally 
observed in chiral magnets ${\rm Cr}{\rm Nb}_3{\rm S}_6$ in an external magnetic field \cite{CM}.

In the chiral limit ($m_{\pi}=0$), Eq.~(\ref{bound}) shows that nuclear matter would be unstable 
against the decay into the space-dependent $\pi^0$ condensation in an infinitesimally small 
magnetic field. In this case, the Hamiltonian density is given by \cite{Thompson:2008qw}
\beq
{\cal H} = \frac{1}{2}({\bm \nabla} \pi^0)^2 - C \mu_{\rm B} {\bm B}_{\rm ex} \cdot {\bm \nabla} \pi^0\,.
\eeq
This is minimized when 
\beq
\label{GS}
\langle {\bm \nabla} \pi^0 \rangle = C \mu_{\rm B} {\bm B}_{\rm ex},
\eeq
and the minimum is
\beq
\langle {\cal H} \rangle = - \frac{1}{2}(C \mu_{\rm B} B_{\rm ex})^2 < 0\,.
\eeq
This is indeed lower than the energy density of nuclear matter for $\mu_{\rm B} \approx m_{\rm N}$.

\section{Effective theory for electromagnetic fields}
Let us consider the low-energy effective theory of \emph{dynamical} electromagnetic 
fields in spatially varying meson condensations. We stress that our argument here 
is independent of the microscopic origin of these phases. As an example, we here consider
the presence of the space-dependent $\pi^0$ condensation and investigate its consequences.

Note that the low-energy effective theory does not respect the full Lorentz symmetry due 
to the presence of the medium, although it does respect the rotational symmetry. So the 
effective Lagrangian can be written, using the gauge-invariant electromagnetic fields, as
\beq
{\cal L}_{\rm EM} = \frac{\epsilon}{2} {\bm E}^2 - \frac{1}{2 \mu} {\bm B}^2\,,
\eeq
with some constants $\epsilon$ and $\mu$ that depend on microscopic details. 
In the usual electromagnetism, $\epsilon$ and $\mu$ are called the permittivity and 
permeability, respectively. We assume that $\epsilon$ and $\mu$ do not depend on 
spatial coordinates and $\epsilon \sim \mu \sim O(1)$.

At low energy, we also need to take into account light meson degrees of freedom 
which couple to electromagnetic fields. In the presence the pion background field, 
we have the following WZW term due to the quantum anomalies \cite{Wess:1971yu, Witten:1983tw}:
\beq
{\cal L}_{\rm WZW} = C \pi^0 {\bm E} \cdot {\bm B}.
% \quad {\rm or} \quad {\cal L}_{\rm WZW} = \frac{\phi}{4\pi^2} {\bm E} \cdot {\bm B}\,, 
\eeq
In the QCD vacuum, this term is related to the decay of a neutron pion to two photons, 
$\pi^0 \rightarrow 2 \gamma$ \cite{Adler, BellJackiw}. The coefficient $C$ is uniquely 
determined by the anomaly matching and is given by Eq.~(\ref{C}) \cite{Wess:1971yu, Witten:1983tw}.

If $\langle \pi^0(\bm x) \rangle$ is a constant in space as usual, this anomalous term is 
just a total derivative and does not affect the dynamics of electromagnetic fields.%
\footnote{We note that, when $\langle \pi^0(\bm x) \rangle \neq 0$, the WZW term above 
leads to anomalous magnetoelectric responses similar to topological insulators \cite{Qi:2008ew}, 
such as the magnetization induced by the electric field, ${\bm M} = C \langle \pi^0(\bm x) \rangle {\bm E}$, 
and the electric polarization induced by the magnetic field, ${\bm P} = C \langle \pi^0(\bm x) \rangle {\bm B}$. 
When $\langle \pi^0(\bm x) \rangle$ is a constant, however, these modifications do not affect 
the equations of motion for electromagnetic fields.} 
However, when $\langle \pi^0({\bm x}) \rangle$ varies spatially, it modifies the equations 
of motion for electromagnetic fields. From the full Lagrangian, 
\beq
\label{L_full}
{\cal L} = {\cal L}_{\rm EM} + {\cal L}_{\rm WZW} + A_{\mu} j^{\mu},
\eeq
with the electric current density $j^{\mu}=(\rho,{\bm j})$, 
the equations of motion for the electromagnetic fields are derived as
\begin{gather}
\label{Gauss}
\epsilon {\bm \nabla} \cdot {\bm E} = \rho - C \langle {\bm \nabla} \pi^0 \rangle \cdot {\bm B}, \\
\label{Ampere}
\frac{1}{\mu} {\bm \nabla} \times {\bm B} = \epsilon \d_t {\bm E} + {\bm j} + C \langle {\bm \nabla} \pi^0 \rangle \times {\bm E}\,.
\end{gather}
The final terms in Eqs.~(\ref{Gauss}) and (\ref{Ampere}) are the additional contributions
to the usual Gauss's law and Amp\`ere's law, respectively, due to the presence of the 
space-dependent $\pi^0$ field. The former correction was already found in Ref.~\cite{Son:2007ny} 
through a different path. On the other hand, the latter correction, which stands for the electric 
current in the direction perpendicular to the electric field, has not yet been appreciated. As 
we will see later, this anomalous Hall current modifies the dispersion relation of photons. 

Note that, if $\langle \pi^0 \rangle$ were time-dependent, $\langle \d_t \pi^0 \rangle$ 
would play the role of the so-called chiral chemical potential and generates the ``chiral 
magnetic effect," ${\bm j} \sim \langle \d_t \pi^0 \rangle {\bm B}$ \cite{Fukushima:2012fg}.

The form of Lagrangian in Eq.~(\ref{L_full}) is equivalent to the axion electrodynamics 
\cite{Wilczek:1987mv}. The axion electrodynamics is also found to emerge at the interface 
between a topological insulator and a normal insulator in condensed matter physics, where 
the ``axion field" $\theta({\bm x})$ varies in space from $\theta=0$ in the normal insulator 
to $\theta=\pi$ in the topological insulator \cite{Qi:2008ew}. In the specific case of the $\pi^0$
domain wall above, the field $\phi(z) = \pi^0(z)/f_{\pi}$ plays the role of the ``axion field," 
which varies from $0$ to $2\pi$ across the wall.

\section{Nonrelativistic photons}
We now consider an electromagnetic wave (or a photon) in the spatially varying meson 
condensation. As an illustration, we consider the simplest case where $\langle {\bm \nabla} \pi^0 \rangle$ 
is homogeneous. Such a configuration is achieved, e.g., by Eq.~(\ref{GS}) in a homogeneous 
external magnetic field \cite{Thompson:2008qw}. Our argument may also be extended to 
periodic meson condensations \cite{CMSL}. We assume the condition of local electric charge 
neutrality, $n=0$, and ${\bm j} = {\bm 0}$.%
\footnote{When the Ohmic current is added, the dispersion relation that we will derive
below has the imaginary part and the electromagnetic wave is damped by dissipation.}
We take the variation of the meson condensation in the $z$ direction,  $\langle \d_z \pi^0 \rangle > 0$, 
and consider the perturbation of electromagnetic fields in the $xy$ plane. In this setup, we have 
$\langle {\bm \nabla} \pi^0 \rangle \cdot {\bm B} = 0$, and so ${\bm \nabla} \cdot {\bm E} = 0$ 
from Eq.~(\ref{Gauss}); the electromagnetic wave is a transverse wave. Below we will 
consider the electromagnetic wave propagating in the positive $z$ direction. 

Combining Eq.~(\ref{Ampere}) and Faraday's law, ${\bm \nabla} \times {\bm E} = - \d_t {\bm B}$, 
and eliminating ${\bm B}$, we obtain a closed equation in terms of ${\bm E}$ as
\beq
\label{wave_eq}
\d_t^2 {\bm E} = v^2 {\bm \nabla}^2 {\bm E} - \frac{C}{\epsilon} \langle {\bm \nabla} \pi^0 \rangle \times \d_t {\bm E}\,,
\eeq
where $v \equiv 1/\sqrt{\epsilon \mu}$ is the velocity of light in medium. Without the meson 
condensation, this reduces to the conventional wave equation for electromagnetic waves.

We then look for the plane-wave solution of the form, 
${\bm E} = {\bm E}_0 e^{-i \omega t + i k z}$, where ${\bm E}_0 = E_x {\bm e}_x + E_y {\bm e}_y$
and $\omega, k>0$. Substituting it into Eq.~(\ref{wave_eq}), we have
\begin{align}
\omega^2 E_x = v^2 k^2 E_x + i \omega \frac{C}{\epsilon} \langle \d_z \pi^0 \rangle E_y\,, \\
\omega^2 E_y = v^2 k^2 E_y - i \omega \frac{C}{\epsilon} \langle \d_z \pi^0 \rangle E_x\,.
\end{align}
From these equations, we get the dispersion relation,
\beq
\label{dispersion}
\omega^2 - v^2 k^2 \pm  \frac{C}{\epsilon} \langle \d_z \pi^0 \rangle \omega = 0\,.
\eeq
Here the signs $\pm$ correspond to the positive and negative helicity states of the photon, 
$E_y = \pm i E_x$, which we denote as $\lambda = \pm 1$.  

For $k \gg C \langle \d_z \pi^0 \rangle$, the dispersion relation (\ref{dispersion}) reduces 
to $\omega = v k$ independently of the helicity. This is the same dispersion as the 
usual electromagnetic wave in medium; the wavelength $1/k$ is so short that the wave 
is not affected by the structure of the meson condensation.

For $k \ll C \langle \d_z \pi^0 \rangle$, on the other hand, the dispersion relation 
(\ref{dispersion}) are different depending on the helicity; the dispersion relations 
for $\lambda = \pm 1$ are given by%
\footnote{For the electromagnetic wave propagating in the negative $z$ direction
($\omega < 0$ and $k > 0$), we have 
$\omega_+ = -m^*\,$ and $\omega_- = -v^2 k^2/m^*$.}
\beq
\label{dispersion_long}
\omega_+ = \frac{v^2 k^2}{m^*}\,, \qquad \omega_- = m^*,
\eeq
respectively, where $m^* = C \langle \d_z \pi^0 \rangle/\epsilon$.
The photon with $\lambda = +1$ is gapless, $\lim_{k \rightarrow 0}\omega_+(k) = 0$,
while the one with $\lambda = -1$ is gapped. Note also that the former has the 
``nonrelativistic" dispersion. In this way, the spatially varying meson condensation 
can distinguish between the two helicity states of photons.

This should be contrasted with Maxwell-Chern-Simons theories (topologically massive 
gauge theories) in two spatial dimensions \cite{Deser:1982vy} and superconductors. 
In these systems, the dispersion relation of photons is $\omega^2 = k^2 + m^2$ with $m$ 
the screening mass, which reduces to a gapped dispersion $\omega = \pm m$ for $k \ll m$ 
and is not the same as Eq.~(\ref{dispersion_long}).

It is known that the usual photons with the dispersion relation $\omega^2 = k^2$ in the 
vacuum can be understood as the gapless Nambu-Goldstone (NG) modes 
\cite{Ferrari:1971at, Brandt:1974jw}. The nonrelativistic gapless photons in Eq.~(\ref{dispersion_long}) 
may be understood as the so-called type-II NG modes \cite{Nielsen:1975hm}%
\footnote{See also Refs.~\cite{Watanabe:2011ec, Watanabe:2012hr, Hidaka:2012ym} 
for the general counting rule of type-I and type-II NG modes.} 
due to the presence of spatially varying meson condensation, in a way similar to 
Ref.~\cite{Kobayashi:2014xua}. This will be reported elsewhere.

\section{Discussion}
In this paper, we argued that the low-energy effective theory for electromagnetic fields in
spatially varying meson condensations is given by the axion electrodynamics. This is analogous 
to the physics at the interface between topological and ordinary insulators. We also found that 
such meson condensations can distinguish between the helicity states of photons, 
and, in particular, one of them is a nonrelativistic gapless photon. Our finding suggests that, if 
such phases are realized in neutrons stars, the electromagnetic properties of neutron stars 
should be modified by the quantum Hall effect.

In this paper, we consider the time-independent (but space-dependent) meson condensation.
If the meson condensation is time-dependent, $\langle \d_t \pi^0 \rangle \neq 0$, 
the chiral magnetic effect of the form ${\bm j} \sim \langle \d_t \pi^0 \rangle {\bm B}$ would 
lead to an unstable mode, which then tends to erase the time dependence of 
$\langle \pi^0 \rangle$ and generates helical magnetic fields. This is an analogue of the 
chiral instability \cite{Akamatsu:2013pjd, Joyce:1997uy} in the meson condensation,
which may have relevance to the origin of the strong magnetic fields in magnetars 
\cite{Ohnishi:2014uea, Yamamoto:2015gzz}.

Finally, we emphasize once again that essentially the same physics should also be realized
at the surface of topological insulators. It would be interesting to experimentally observe 
the nonrelativistic gapless photons, which is a novel signature of the ``axion field" 
$\theta$, in condensed matter systems.

\acknowledgments
The author thanks Y. Hidaka for useful comments. This work was supported, in part, 
by JSPS KAKENHI Grants No.~26887032 and MEXT-Supported Program for the 
Strategic Research Foundation at Private Universities, ``Topological Science'' 
(Grant No.~S1511006). 

{\it Note added.}---After this work was being completed, we learned that a closely 
related observation was found in Ref.~\cite{Ferrer:2015iop}, where the axion 
electrodynamics emerges in the dual chiral density wave (DCDW). While 
Ref.~\cite{Ferrer:2015iop} uses an NJL-type model and considers the DCDW, our 
argument is based on the low-energy effective theory and is model-independent; 
our assumption is rather the presence of the gradient of a meson field alone, and 
it does not necessarily require the CDW itself. Also, the nonrelativistic gapless 
photons found in this paper were not discussed in Ref.~\cite{Ferrer:2015iop}.

\end{document}